\documentclass[journal]{IEEEtran}

\ifCLASSINFOpdf
  \usepackage[pdftex]{graphicx}

\else

  \usepackage[dvips]{graphicx}

\fi
\usepackage{subfig}
\usepackage{xcolor}
\usepackage{multirow}
\usepackage{booktabs}
\usepackage{pifont}
\usepackage{cite}
\usepackage{pdfcomment}
\usepackage{amsmath}
\usepackage{amsthm}
\usepackage{amssymb}
\usepackage{soul}
\usepackage{caption}
\usepackage{cleveref}

\begin{document}
%
\title{U6G XL-MIMO Radiomap Prediction: Multi-Config Dataset and Beam Map Approach}

\author{Xiaojie Li,~\IEEEmembership{Student Member,~IEEE,}
Yu Han,~\IEEEmembership{Member,~IEEE,}
Zhizheng Lu,~\IEEEmembership{Student Member,~IEEE,}

Shi Jin,~\IEEEmembership{Fellow,~IEEE,}
and Chao-Kai Wen,~\IEEEmembership{Fellow,~IEEE}

\thanks{Xiaojie Li, Yu Han, Zhizheng Lu, and Shi Jin are with the School of Information Science and Engineering, Southeast University, Nanjing 210096,
China (e-mail: \{xiaojieli, hanyu, luzz and jinshi\}@seu.edu.cn).

Chao-Kai Wen is with the Institute of Communications Engineering,
National Sun Yat-sen University, Kaohsiung 80424, Taiwan (e-mail:
chaokai.wen@mail.nsysu.edu.tw).

}}

\markboth{IEEE Transactions on Wireless Communications}%
{Shell \MakeLowercase{\textit{et al.}}: Bare Demo of IEEEtran.cls for IEEE Journals}

\maketitle

\begin{abstract}
The upper 6 GHz (U6G) band with extremely large-scale multiple-input multiple-output (XL-MIMO) is a key enabler for sixth-generation wireless systems, yet intelligent radiomap prediction for such systems remains challenging. Existing datasets support only small-scale arrays (up to $8\times8$) with predominantly isotropic antennas, far from the 1024-element directional arrays envisioned for 6G. Moreover, current methods encode array configurations as scalar parameters, forcing neural networks to extrapolate array-specific radiation patterns, which fails when predicting radiomaps for configurations absent from training data. To jointly address data scarcity and generalization limitations, this paper advances XL-MIMO radiomap prediction from three aspects. To overcome data limitations, we construct the first XL-MIMO radiomap dataset containing 78400 radiomaps across 800 urban scenes, five frequency bands (1.8–6.7 GHz), and nine array configurations up to $32\times32$ uniform planar arrays with directional elements. To enable systematic evaluation, we establish a comprehensive benchmark framework covering practical scenarios from coverage estimation without field measurements to generalization across unseen configurations and environments. To enable generalization to arbitrary beam configurations without retraining, we propose the beam map, a physics-informed spatial feature that analytically computes array-specific coverage patterns. By decoupling deterministic array radiation from data-learned multipath propagation, beam maps shift generalization from neural network extrapolation to physics-based computation. Integrating beam maps into existing architectures reduces mean absolute error by up to 60.0\% when generalizing to unseen configurations and up to 50.5\% when transferring to unseen environments. The complete dataset and code are publicly available at {https://lxj321.github.io/MulticonfigRadiomapDataset/}.

\end{abstract}


\IEEEpeerreviewmaketitle

\section{Introduction}\label{sec:intro}


\IEEEPARstart{T}{he} evolution toward 5G-Advanced (5G-A) and sixth-generation (6G) wireless systems demands substantial new spectrum to support enhanced network capacity \cite{10835238}. To address this demand, the World Radiocommunication Conference 2023 (WRC-23) allocated up to 700 MHz of spectrum in the upper 6 GHz (U6G) band (6.425--7.125 GHz) for International Mobile Telecommunications (IMT) services \cite{ghosh2023world}. However, the U6G band exhibits approximately 5.6~dB higher free-space path loss than conventional sub-6~GHz bands (e.g., 3.5~GHz) for the same propagation distance, posing coverage challenges in urban deployments~\cite{miao2023sub}.

Extremely large-scale multiple-input multiple-output (XL-MIMO) has emerged as a key enabling technology to address this elevated path loss \cite{lu2024tutorial,11050910}. By deploying large apertures such as $32 \times 32$ uniform planar arrays (UPAs) with 1024 antenna elements, XL-MIMO systems achieve substantial beamforming gains that compensate for the increased path loss at higher frequencies \cite{11017428}. While physical-layer research has extensively studied XL-MIMO link-level design, including channel modeling \cite{ding2025far}, beam management \cite{li2025codebook}, and hybrid precoding \cite{10944717}, network-level spatial coverage prediction remains underexplored for such large-scale array systems.

Such network-level coverage prediction, known as \textit{radiomap prediction}, aims to infer received signal strength distribution across a geographical area from transmitter configuration and environmental features. Accurate radiomap prediction plays a critical role in multiple network applications: it enables coverage estimation for network planning prior to base station deployment \cite{10877906}, facilitates interference-aware resource allocation in dense networks \cite{10556774}, and supports fingerprint-based localization services \cite{yapar2023real}. For XL-MIMO systems, radiomap prediction becomes particularly challenging due to the vast configuration space spanning carrier frequencies, array architectures, and beam directions, all of which significantly influence spatial coverage patterns.

Conventional radiomap prediction relies on model-based approaches, which can be broadly divided into two categories: analytical/empirical propagation models and ray tracing methods. The former, such as the Friis free-space model and 3GPP urban macro (UMa) model \cite{etsi2020study}, construct radiomaps by computing path loss as functions of carrier frequency, link distance, and antenna height at each spatial location. While computationally efficient, these models lack consideration for specific urban geometries, such as building layouts and terrain variations, limiting radiomap accuracy in complex environments \cite{zeng2024tutorial}. Ray tracing offers higher-fidelity radiomap generation by simulating multipath propagation through ray emission, reflection, and diffraction analysis \cite{hoydis2022sionna}. However, ray tracing requires accurate 3D environment modeling and incurs substantial computational overhead \cite{de2021convergent}. For XL-MIMO systems, the computational burden further escalates: generating a single radiomap requires evaluating channel responses for all array elements and computing beamforming gains across the entire coverage area, making exhaustive ray tracing impractical for large-scale network planning that involves numerous candidate configurations. These limitations motivate the exploration of data-driven approaches for radiomap prediction.

\begin{table*}[t]
\caption{Comparison with Existing Radiomap Datasets}
\label{tab:dataset_comparison}
\centering
\begin{tabular}{lcccccc}
\toprule
\textbf{Dataset} & \textbf{Area (m$^2$)} & \textbf{Scenes} & \textbf{Multi-band} & \textbf{MIMO} & \textbf{Antenna} & \textbf{Height Map} \\
\midrule
\textbf{Ours} & $1280\times1280$ & 800 & \checkmark & $32\times32$ UPA & directional & \checkmark \\
DeepMIMO v4.0b \cite{deepmimo} & N/A & 152 & \checkmark & $8\times8$ UPA & isotropic & \checkmark \\
BeamCKM \cite{beamckm} & $256\times256$ & 100 & $\times$ & $8$ ULA & isotropic & \checkmark \\
UrbanRadio3D \cite{radiodiff3d} & $256\times256$ & 701 & $\times$ & $\times$ & isotropic & \checkmark \\
CKMImageNet \cite{wu2024ckmimagenet,wu2025ckmimagenet} & $1280\times1280$ & 42 & $\times$ & $\times$ & isotropic & \checkmark \\
SpectrumNet \cite{spectrumnet} & $1280\times1280$ & 764 & \checkmark & $\times$ & isotropic & $\times$  \\
RadioGAT \cite{radiogat} & $1000\times1000$ & 10 & \checkmark & $\times$ & isotropic & $\times$ \\
RMDirectionalBerlin \cite{rmdirectionalberlin} & $256\times256$ & 424 & $\times$ & $\times$ & directional & $\times$ \\
RadioMapSeer \cite{radiounet} & $256\times256$ & 701 & $\times$ & $\times$ & isotropic & $\times$ \\
\bottomrule
\end{tabular}
\end{table*}

Data-driven methods offer an alternative paradigm that exploits statistical characteristics of radiomap data to predict signal strength distributions, reducing the reliance on explicit propagation modeling. Early efforts employed traditional machine learning techniques, including Gaussian process regression \cite{b21}, matrix completion \cite{b22}, and Kriging interpolation \cite{b23}, for radiomap reconstruction from limited spatial observations. More recently, deep learning methods have demonstrated superior capability in capturing complex propagation patterns. Representative approaches include RadioUNet \cite{radiounet}, RME-GAN \cite{rmegan}, RadioGAT \cite{radiogat}, RadioDiff \cite{radiodiff}, and UniRM \cite{unirm}, achieving promising accuracy by learning to predict spatial signal strength from environmental features such as building height maps. However, these methods universally assume omnidirectional (isotropic) transmitters, which fundamentally differ from practical deployments utilizing directional antenna elements.

Recent works have begun addressing directional transmission. UNetDCN \cite{rmdirectionalberlin} encodes single-element radiation patterns as 2D gain projections onto the coverage plane, while BeamCKM \cite{beamckm} employs beam index embedding to implicitly associate beam directions with coverage patterns. Despite these advances, both approaches exhibit limitations for XL-MIMO systems: the 2D gain projection cannot capture the combined effect of array architecture and beamforming, whereas beam index embedding requires retraining for different array configurations, limiting scalability to the vast XL-MIMO configuration space. Furthermore, existing radiomap datasets constrain progress in this direction: as shown in Table~\ref{tab:dataset_comparison}, the largest MIMO-capable dataset supports only $8 \times 8$ UPAs \cite{deepmimo}, far below the 1024-element arrays envisioned for 6G systems. Moreover, most datasets adopt isotropic antenna models that cannot capture directional beamforming effects, and several lack building height information essential for learning environment-propagation relationships.

The aforementioned limitations in methods and datasets are compounded by the diverse requirements of practical deployment scenarios. Network planners need \textit{blind prediction} to estimate coverage for planned base stations before field measurements are available, and \textit{sparse reconstruction} to complete radiomaps from limited samples collected through drive tests or crowdsourcing. More critically, deployed models must generalize across the vast XL-MIMO configuration space without retraining\cite{11268973}. In practical network planning, operators evaluate coverage across diverse settings: different carrier frequencies for capacity planning, various array sizes for cost-performance tradeoffs, and multiple beam directions for sector optimization. If prediction models require retraining for each new configuration, the associated data collection and computational costs become prohibitive. Therefore, \textit{cross-configuration generalization}, the ability to predict radiomaps for array configurations absent from training data, is essential for practical deployment. Similarly, \textit{cross-environment generalization} to unseen geographical areas is required when extending coverage prediction to new deployment regions.

Existing works have explored these generalization dimensions in isolation. For cross-environment generalization, RadioUNet \cite{radiounet}, RME-GAN \cite{rmegan}, and RadioDiff \cite{radiodiff} train on diverse urban scenes to enable transfer to unseen geographical areas. For cross-configuration generalization, RadioGAT \cite{radiogat} and UniRM \cite{unirm} evaluate cross-frequency transferability; UNetDCN \cite{rmdirectionalberlin} and the ICASSP 2025 Challenge \cite{bakirtzis2025indoorpathlossradiomap} test cross-element-pattern generalization for single antennas; BeamCKM \cite{beamckm} supports cross-beam-direction prediction through beam index conditioning. However, these efforts adopt inconsistent evaluation protocols, cover different subsets of the configuration space, and use different dataset partitioning strategies, preventing systematic comparison of generalization capabilities across methods.

These observations reveal that current research cannot adequately support XL-MIMO radiomap prediction. From the data perspective, existing datasets are limited to small arrays (up to $8 \times 8$ UPAs) with predominantly isotropic antenna models, far from the 1024-element directional arrays envisioned for 6G systems. The computational complexity of generating XL-MIMO radiomaps, arising from large-scale channel matrices and extensive spatial coverage grids across multiple configurations, has hindered large-scale dataset construction. From the evaluation perspective, current benchmarks remain fragmented across individual generalization dimensions with inconsistent protocols, preventing systematic assessment of prediction accuracy and generalization capabilities. From the methodology perspective, existing encoding schemes for directional transmission cannot scale to the XL-MIMO configuration space: 2D gain projections fail to capture the combined effect of array architecture and beamforming, while beam index embeddings require retraining for each new configuration.

To address these challenges, we make three contributions that jointly establish the foundation for XL-MIMO radiomap prediction research:

\begin{itemize}
\item \textbf{XL-MIMO Radiomap Dataset.} We construct the first large-scale radiomap dataset supporting XL-MIMO research, containing 78400 radiomaps across 800 urban scenes, five frequency bands (1.8--6.7 GHz), and nine array configurations from $2\times2$ to $32\times32$ UPAs with directional antenna elements. By leveraging GPU-accelerated ray tracing, we overcome the computational barrier that has previously precluded large-scale XL-MIMO radiomap dataset construction.

\item \textbf{Systematic Evaluation Framework.} We establish standardized benchmarks comprising three complementary tasks: blind prediction for coverage estimation without field measurements, sparse reconstruction from limited spatial samples, and cross-distribution generalization to unseen configurations or environments. This framework enables systematic comparison of prediction accuracy and generalization capabilities across methods.

\item \textbf{Beam Map Representation.} We propose the beam map, a physics-informed spatial feature that analytically computes array-specific line-of-sight (LoS) coverage patterns from configuration parameters, incorporating element radiation patterns, array geometry, beamforming vectors, and free-space propagation characteristics. While environment-dependent multipath propagation is learned from data, array-dependent LoS radiation can be computed deterministically. This decoupling shifts cross-configuration generalization from data-driven extrapolation to physics-based computation. Integrating beam maps into existing architectures reduces mean absolute error by up to 51.5\% for blind prediction (RME-GAN) and 60.0\% for cross-configuration generalization (RadioUNet).

\end{itemize}

The remainder of this paper is organized as follows. Section~\ref{sec:system_model} presents the system model and problem formulation. Section~\ref{sec:dataset} describes the dataset construction methodology. Section~\ref{sec:task_taxonomy} defines the evaluation task taxonomy. Section~\ref{sec:beam_map} introduces the beam map representation. Section~\ref{sec:experiment} reports experimental results, and Section~\ref{sec:conclusion} concludes the paper.

\textit{Notation:} Boldface uppercase and lowercase letters denote matrices and vectors, respectively. $(\cdot)^T$, $(\cdot)^H$, and $(\cdot)^*$ represent transpose, conjugate transpose, and complex conjugate. $|\cdot|$ and $\|\cdot\|$ denote absolute value and Euclidean norm. $\mathbb{C}^{m \times n}$ and $\mathbb{R}^{m \times n}$ denote complex and real matrix spaces.

\section{System Model and Problem Formulation}\label{sec:system_model}

This section establishes the mathematical framework for XL-MIMO radiomap modeling. Section~\ref{sec:radiomap_definition} defines the antenna array architecture and radiomap representation. Section~\ref{sec:radiomap_modeling} develops the physical model, integrating directional antenna patterns, multipath channels, beamforming and radiomap calculation.

\subsection{XL-MIMO Radiomap Definition}\label{sec:radiomap_definition}

Consider an XL-MIMO base station equipped with a transmit array configured as an $N_{\text{rows}} \times N_{\text{cols}}$ UPA, comprising a total of $N = N_{\text{rows}} \times N_{\text{cols}}$ antenna elements. The inter-element spacing is set to $d = 0.5\lambda$, where $\lambda = c/f_c$ represents the carrier wavelength, with $c = 3 \times 10^8 \, \text{m/s}$ denoting the speed of light and $f_c$ representing the carrier frequency. 

\begin{figure}
    \centering
    \includegraphics[width=0.8\linewidth]{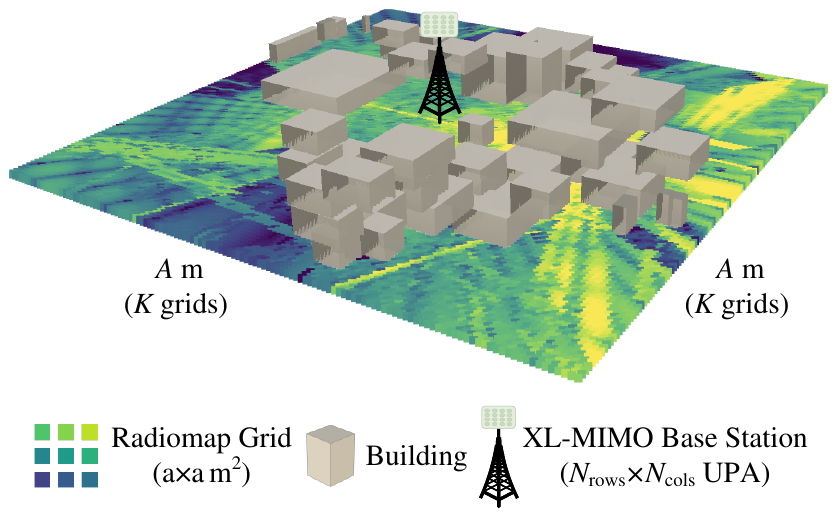}
\caption{System model of XL-MIMO radiomap prediction, where the radiomap characterizes spatial signal strength distribution across a $K\times K$ grid observation plane with color indicating received power level (yellow: high, dark blue: low).}
    \label{fig:system_model}
\end{figure}

Without loss of generality, we assume that the UPA is deployed at the height $h^{(\text{height})}_{\text{BS}}$. We further consider a two-dimensional observation plane of size $A \times A$ $\mathrm{m}^2$ at height $h^{(\text{height})}_{\text{Map}}$, centered at the horizontal projection of the UPA, as illustrated in Fig.~\ref{fig:system_model}. This plane is discretized into $K \times K$ uniform grids, each with area $a \times a$ $\mathrm{m}^2$, collectively denoted as $\mathbf{R} = [\mathcal{R}_1, \ldots, \mathcal{R}_k, \ldots, \mathcal{R}_{K^2}]$. Following \cite{zhang2023physics,11268973}, the coverage at each grid $\mathcal{R}_k$ is characterized by the measured average Synchronization Signal Block Reference Signal Received Power (SSB-RSRP) $P_{r,k}$. \footnote{Note that the final coverage of a base station is determined by the maximum SSB-RSRP across time-division scanned SSB beams in different grids, known as SS-RSRP. However, we focus on SSB-RSRP for two reasons: (1) each SSB beam corresponds to a specific precoding matrix, which facilitates subsequent modeling, and (2) user equipment reports the RSRP of each SSB beam during measurements. With SSB-RSRP-level radiomap characterization capability, SS-RSRP-level radiomaps can be derived by aggregating multiple SSB-RSRP radiomaps. We note that standard SSB-RSRP is defined as the linear average power within the 240-subcarrier SSB bandwidth; our model adopts a single-carrier approximation, which yields negligible deviation for path loss characterization and serves as a reasonable baseline for XL-MIMO radiomap research.} Consequently, the radiomap corresponding to such an XL-MIMO base station is defined as $\mathbf{P}_r = [P_{r,1}, \ldots, P_{r,k}, \ldots, P_{r,K^2}]$. Although the radiomap depends on transmit power through received signal strength, the spatial coverage pattern itself, which is determined by beamforming gains and multipath effects, is independent of transmit power.

\subsection{XL-MIMO Radiomap Modeling}\label{sec:radiomap_modeling}
In this subsection, we develop a comprehensive physical model for the XL-MIMO radiomap by systematically integrating multiple propagation mechanisms. We proceed as follows: First, we characterize the radiation pattern of individual antenna elements. Second, we model the multipath channel with element-specific propagation characteristics. Third, we formulate the SSB beam precoding vectors. Fourth, we derive the received power at an arbitrary grid $\mathcal{R}_k$. Finally, we discuss power normalization conventions that ensure model generality across transmit power configurations.

\subsubsection{Antenna Element Radiation Pattern}
In XL-MIMO systems, the radiation pattern of each antenna element follows the 3GPP TR 38.901 specification \cite{etsi2020study}. Let $\theta\in[0, \pi]$ denote the zenith angle measured from the array boresight and $\phi \in [-\pi, \pi]$ denote the azimuth angle. The composite element gain in dB is given by
\begin{equation}
    G(\theta, \phi) = G_{E,\text{max}} - \min\left\{A_{V}(\theta) + A_{H}(\phi),\, A_{\text{max}}\right\},\label{eq1}
\end{equation}
where $G_{E,\text{max}} = 8$~dBi is the maximum directional gain, $A_{\text{max}} = 30$~dB is the front-to-back attenuation, and the vertical/horizontal attenuation functions are
\begin{align}
    A_{V}(\theta) &= \min\left\{12\left(\frac{\theta - \pi/2}{\theta_{3\text{dB}}}\right)^2,\, \text{SLA}_V\right\}, \label{eq1a}\\
    A_{H}(\phi) &= \min\left\{12\left(\frac{\phi}{\phi_{3\text{dB}}}\right)^2,\, A_{\text{max}}\right\}, \label{eq1b}
\end{align}with half-power beamwidth $\theta_{3\text{dB}} = \phi_{3\text{dB}} = 65^{\circ}$ and vertical side-lobe level $\text{SLA}_V = 30$~dB. For notational convenience, we denote the linear-scale gain as $G_t^{(n)}(\theta, \phi) = 10^{G(\theta,\phi)/10}$ for subsequent channel modeling.

\subsubsection{Multipath Channel Modeling}
Due to the physically large aperture of XL-MIMO arrays, different antenna elements may experience spatially distinct propagation paths when communicating with a given grid $\mathcal{R}_k$.\footnote{For analytical tractability, we assume the receive antenna is located at the center of grid $\mathcal{R}_k$.} Assuming a maximum of $N_p$ multipaths for each antenna element, we model the channel from the UPA to grid $\mathcal{R}_k$ as a two-dimensional matrix
\begin{equation}
    \mathbf{H}_k = [h_{n,p,k}] \in \mathbb{C}^{N \times N_p}, \quad n = 1, \ldots, N, \quad p = 1, \ldots, N_p\label{eq2}
\end{equation}
where $h_{n,p,k}$ represents the complex channel coefficient of the $p$-th propagation path associated with the $n$-th antenna element. 

Specifically, incorporating both the channel propagation characteristics and the directional antenna gain, $h_{n,p,k}$ is expressed as
\begin{equation}
    h_{n,p,k} = \sqrt{G_t^{(n)}(\theta_{n,p,k}, \phi_{n,p,k})} \cdot \frac{e^{j 2\pi  r_{n,p,k}/\lambda}}{r_{n,p,k}}\cdot \Gamma_{n,p,k},\label{eq3}
\end{equation}
where $G_t^{(n)}(\theta_{n,p,k}, \phi_{n,p,k})$ is the antenna gain of the $n$-th element in the direction of the $p$-th path, characterized by zenith angle $\theta_{n,p,k}$ and azimuth angle $\phi_{n,p,k}$. The term $r_{n,p,k}$ denotes the propagation distance of the $p$-th path from antenna element $n$ to grid $k$, and the path interaction coefficient $\Gamma_{n,p,k} \in \mathbb{C}$ captures the cumulative complex-valued attenuation and phase shifts arising from all reflection and diffraction interactions along the $p$-th path:

\begin{equation}
\Gamma_{n,p,k} = \prod_{m=1}^{M_{n,p,k}} R_{n,p,k,m} \cdot \prod_{l=1}^{L_{n,p,k}} D_{n,p,k,l},\label{eq4}
\end{equation}
where $M_{n,p,k}$ denotes the number of reflections encountered along the $p$-th path from the $n$-th antenna element to grid $k$, with $R_{n,p,k,m} \in \mathbb{C}$ representing the complex reflection coefficient of the $m$-th reflection point; $L_{n,p,k}$ denotes the number of diffractions along this path, with $D_{n,p,k,l} \in \mathbb{C}$ representing the complex diffraction coefficient of the $l$-th diffraction point.

The reflection coefficient $R_{n,p,k,m}$ is determined by the material relative permittivity $\epsilon_r$ and the incident angle $\theta_i$ according to the Fresnel equations. For vertical polarization adopted in this work\footnote{Vertical polarization is adopted as the baseline single-polarization configuration in 3GPP TR 38.901~\cite{etsi2020study}, which is widely used in system-level simulations. The methodology extends directly to dual-polarized ($\pm 45^{\circ}$) configurations by incorporating both polarization components.}, it is given by
\begin{equation}
R_{n,p,k,m} = \frac{\sqrt{\epsilon_r - \sin^2\theta_i} - \epsilon_r \cos\theta_i}{\sqrt{\epsilon_r - \sin^2\theta_i} + \epsilon_r \cos\theta_i}.\label{eq4a}
\end{equation}
The diffraction coefficient $D_{n,p,k,l}$ is computed using the Uniform Theory of Diffraction (UTD) \cite{balanis2016antenna}.

For notational convenience in subsequent analysis, we define the equivalent power path gain as $\gamma_{n,p,k} = |\Gamma_{n,p,k}|^2$, representing the power attenuation ratio due to all interactions along the propagation path.

\subsubsection{SSB Beam Precoding}
For a given SSB beam, the precoding vector is denoted as $\mathbf{w}=[w_1,\ldots,w_n,\ldots,w_N]^T \in \mathbb{C}^{N \times 1}$, satisfying the power normalization constraint $\|\mathbf{w}\|^2 = 1$ to ensure conservation of total transmit power. Following 5G NR specifications, we adopt conventional far-field beam steering with uniformly sampled azimuth directions. Although near-field-optimized precoding designs have been proposed \cite{li2025codebook,10944717}, far-field codebook is adopted to maintain consistency with current commercial SSB beam sweeping, ensuring direct applicability to practical network planning. The precoding coefficient for the $n$-th antenna element is given by
\begin{equation}
w_n = \frac{1}{\sqrt{N}} \cdot  e^{j 2\pi \cdot \frac{d}{\lambda} \cdot (x_n \sin \theta_{\text{beam}} \cos \phi_{\text{beam}} + y_n \sin \theta_{\text{beam}} \sin\phi_{\text{beam}})},\label{eq5}
\end{equation}
where $(x_n, y_n)$ represents the relative coordinates of the $n$-th antenna element with respect to the array center (normalized by inter-element spacing $d$), and $(\theta_{\text{beam}}, \phi_{\text{beam}})$ denote the beam steering direction. The zenith angle $\theta_{\text{beam}}$ is measured from the array boresight (perpendicular to the UPA plane), and the azimuth angle $\phi_{\text{beam}}$ specifies the horizontal steering direction, where $\phi_{\text{beam}} < 0$ corresponds to counter-clockwise rotation and $\phi_{\text{beam}} > 0$ to clockwise rotation when viewed from above. In this work, we focus on azimuth-domain beam steering with $\theta_{\text{beam}} = 90^{\circ}$ (horizontal plane), such that beam directions in Table~\ref{tab:config_space} refer exclusively to $\phi_{\text{beam}}$.

\subsubsection{Received Power Calculation}
Under the control of precoding vector $\mathbf{w}$, the electromagnetic fields radiated from all antenna elements coherently superpose at grid $\mathcal{R}_k$. We assume that grid $\mathcal{R}_k$ is equipped with an isotropic receive antenna characterized by unit gain $G_r = 1$ (0~dBi) and perfect polarization matching. Under this assumption, the receive antenna exhibits uniform response to signals arriving from all directions, allowing the contributions from multiple propagation paths to be aggregated with equal weighting.

The resulting power density of the XL-MIMO array at grid $\mathcal{R}_k$ is given by
\begin{equation}
S_{\text{array}}(\mathcal{R}_k) = \frac{P_t}{4\pi} \left| \mathbf{w}^H \mathbf{H}_k \mathbf{1}_{N_p} \right|^2,\label{eq6}
\end{equation}
where $P_t$ denotes the total transmit power, and $\mathbf{1}_{N_p} \in \mathbb{R}^{N_p \times 1}$ is a vector of ones that aggregates the contributions from all $N_p$ propagation paths.

According to antenna theory, the effective aperture of an isotropic antenna is
\begin{equation}
A_e = \frac{\lambda^2 G_r}{4\pi} = \frac{\lambda^2}{4\pi}.\label{eq7}
\end{equation}

The received power at grid $\mathcal{R}_k$ is obtained by integrating the incident power density over the effective aperture:
\begin{equation}
P_{r,k} = A_e \cdot S_{\text{array}}(\mathcal{R}_k) = \frac{\lambda^2}{(4\pi)^2} P_t \left| \mathbf{w}^H \mathbf{H}_k \mathbf{1}_{N_p} \right|^2.\label{eq8}
\end{equation}

\subsubsection{Power Normalization and Generality}\label{sec:power_normalization}
We normalize transmit power to $P_t = 1$~mW (0~dBm). The radiomap $\mathbf{P}_r$ then directly represents path loss and beamforming gain. For other transmit powers $P_t'$, the received power is $P_{r,k}'[\text{dBm}] = P_{r,k}[\text{dBm}] + (P_t'[\text{dBm}] - P_t[\text{dBm}])$.

By computing $P_{r,k}$ for all grids in the observation plane, we obtain the complete XL-MIMO radiomap, 
\begin{equation}
    \mathbf{P}_r = \begin{bmatrix}
P_{r,1} & \cdots & P_{r,K} \\
\vdots & \ddots & \vdots \\
P_{r,K^2-K+1} & \cdots & P_{r,K^2}
\end{bmatrix},\label{eq12}
\end{equation}
which provides a comprehensive characterization of the spatial coverage performance of the XL-MIMO base station.

\begin{figure*}[t]
    \centering
    \includegraphics[width=0.75\linewidth]{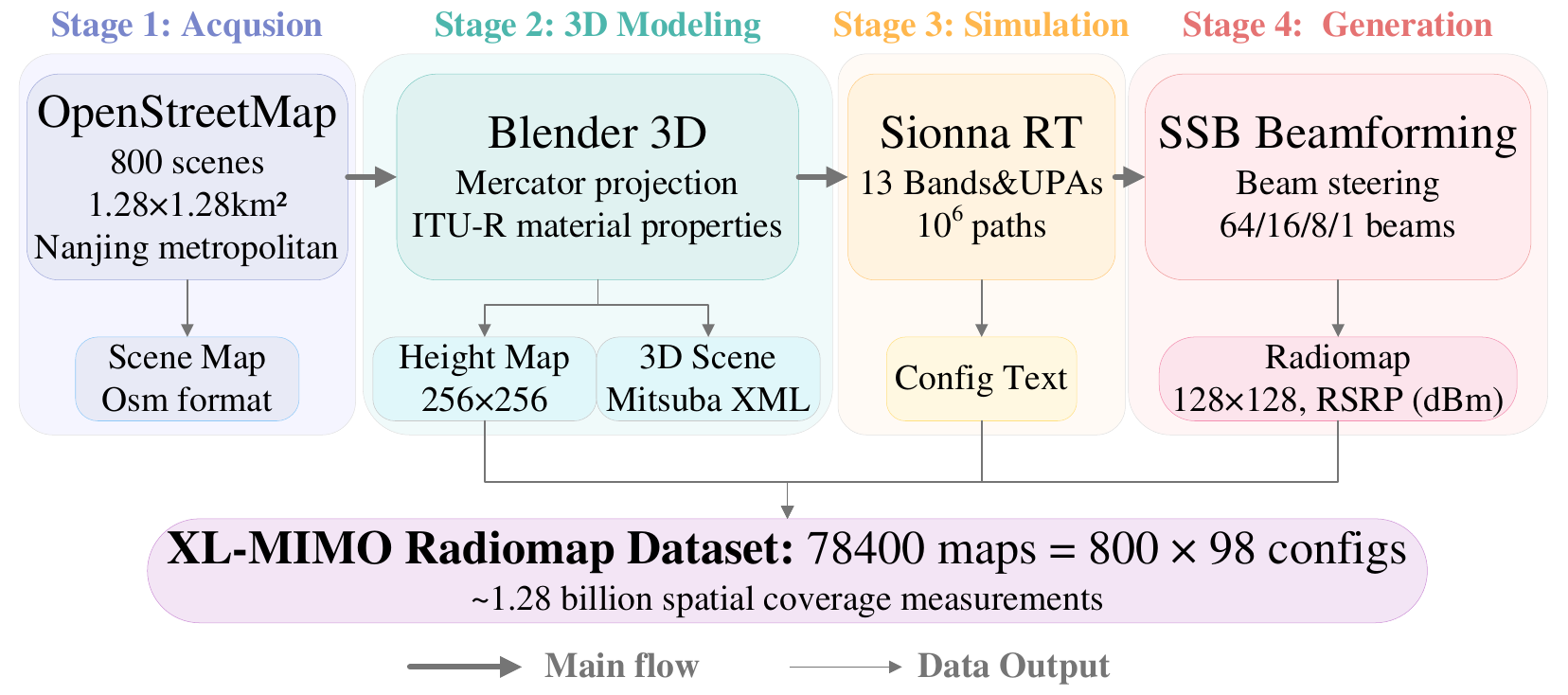}
\caption{XL-MIMO radiomap dataset generation pipeline comprising geographical acquisition, 3D modeling, ray-tracing simulation, and SSB beamforming stages.}
    \label{fig:dataset_pipeline}
\end{figure*}

\section{XL-MIMO Radiomap Dataset Construction}\label{sec:dataset}
Building upon the physical modeling framework established in Section~\ref{sec:system_model}, this section presents the construction methodology for a comprehensive XL-MIMO radiomap dataset. As shown in Fig.~\ref{fig:dataset_pipeline}, the pipeline encompasses: (1) geographical data acquisition from OpenStreetMap (OSM)\footnote{\url{https://www.openstreetmap.org/}} covering 800 urban scenes in the Nanjing metropolitan area, (2) 3D environment modeling via Blender~\cite{soni2023review} with ITU-R standardized material properties, (3) multipath channel computation using Sionna's GPU-accelerated ray tracing engine with adaptive receiver batching~\cite{hoydis2022sionna}, and (4) SSB beamforming and multi-configuration radiomap generation with up to 64 beam directions per array across five frequency bands.

\subsection{Urban Scene Acquisition and Three-Dimensional Modeling}\label{sec:scene_acquisition}
To capture diverse propagation characteristics, we acquired 800 urban scenes from the Nanjing metropolitan area ($31.14^{\circ}N-32.37^{\circ}N, 118.22^{\circ}E-119.14^{\circ}E$) through the OSM Overpass API. Each sample encompasses a $1.28 \times 1.28$~km$^2$ region, dimensioned to adequately represent the extended coverage footprint characteristic of XL-MIMO base stations. The selection process incorporated density-based filtering, retaining regions containing a minimum of five buildings to ensure sufficient environmental complexity for meaningful propagation analysis.

We transformed the acquired vector-format geographical data into ray-tracing-compatible volumetric representations using Blender\cite{soni2023review}, an open-source three-dimensional modeling platform. Geographic coordinates were transformed to local Cartesian frames via Web Mercator projection (EPSG:3857)~\cite{stefanakis2017web}. Topological errors in the OSM data, such as self-intersecting polygons and unclosed boundaries, were corrected programmatically. Building heights were derived from "building:levels" metadata assuming 5~m per level, with a default height of 20~m when metadata was unavailable.

Material properties follow ITU-R P.2040~\cite{sector2024effects}: building surfaces are assigned the dielectric properties of marble, roofs are metallic, and the ground is concrete, with electromagnetic parameters (relative permittivity and conductivity) determined for each carrier frequency. The resulting three-dimensional scenes were exported to Mitsuba XML format compatible with Sionna's ray-tracing coordinate system\cite{hoydis2022sionna}, preserving both geometric fidelity and electromagnetic material specifications essential for subsequent propagation simulation.

In addition to three-dimensional scene models for ray tracing, we generate building height matrices to serve as environmental features for subsequent prediction tasks. The $1280 \times 1280$~m$^2$ area is rasterized into a $256 \times 256$ grid (5~m resolution), with each cell recording the maximum building height.

\subsection{Ray-Tracing Simulation and SSB Beamforming}\label{sec:propagation_simulation}
Building upon the 3D scene models from Section~\ref{sec:scene_acquisition}, this section describes the remaining two stages of the dataset generation pipeline (Fig.~\ref{fig:dataset_pipeline}): multipath channel computation using Sionna's GPU-accelerated ray-tracing engine~\cite{hoydis2022sionna}, and SSB beamforming to obtain radiomaps. The transmitter is positioned at coordinates $(0, 0, 40)$~m, representative of typical macro base station deployments. The observation plane spans $1280 \times 1280$~m$^2$ discretized into a $128 \times 128$ uniform grid, with each $10 \times 10$~m$^2$ grid containing a receiver at height $h^{(\text{height})}_{\text{Map}} = 1.5$~m, yielding 16384 spatial samples per radiomap.

\begin{table}[t]
\centering
\caption{Multi-Dimensional Configuration Parameter Space}
\label{tab:config_space}
\begin{tabular}{ccccc}  
\toprule
\textbf{Band} & \textbf{Antenna (UPA)} & \textbf{Beams} & \textbf{Sweeping} & \textbf{Interval} \\   
\midrule
\multirow{4}{*}{6.7~GHz} & $32\times32$ & 64 & $[-32^{\circ},31^{\circ}]$ & $1^{\circ}$ \\
 & $16\times16$ & 16 & $[-28^{\circ},24.5^{\circ}]$ & $3.5^{\circ}$ \\
 & $8\times8$ & 8 & $[-28^{\circ},21^{\circ}]$ & $7^{\circ}$ \\
 & $16\times32$, $8\times16$ & 1 & $0^{\circ}$ & -- \\
4.9 GHz & $8\times16$, $8\times8$ & 1 & $0^{\circ}$ & -- \\
3.5 GHz & $8\times8$ & 1 & $0^{\circ}$ & -- \\
2.6 GHz & $8\times8$, $4\times8$ & 1 & $0^{\circ}$ & -- \\
1.8 GHz & $4\times4$, $2\times4$, $2\times2$ & 1 & $0^{\circ}$ & -- \\
\bottomrule
\end{tabular}
\end{table}
\begin{figure}[t]
\centering
\includegraphics[width=0.9\linewidth]{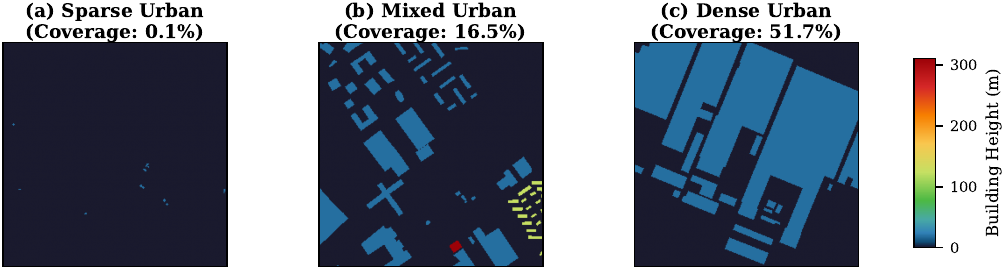}
\caption{Geographic diversity of the proposed dataset: (a) sparse urban (0.1\% building coverage), (b) mixed urban (16.5\%), and (c) dense urban (51.7\%) scenes.}
\label{fig:scene_diversity}
\end{figure}

\begin{figure*}[t]
\centering
\includegraphics[width=0.84\linewidth]{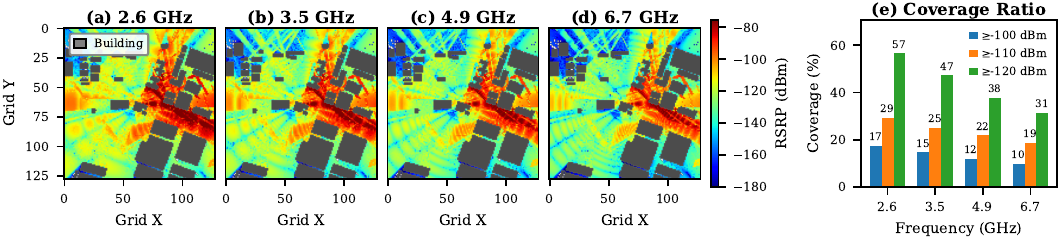}
\caption{Frequency-dependent coverage characteristics for an $8\times8$ UPA: (a)--(d) radiomaps across four carrier frequencies (2.6--6.7~GHz), and (e) coverage ratio at three RSRP thresholds.}
\label{fig:freq_comparison}
\end{figure*}

\begin{figure*}[t]
\centering
\includegraphics[width=0.84\linewidth]{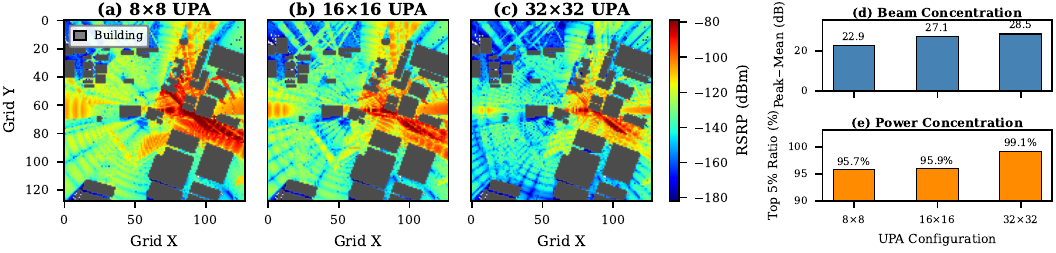}
\caption{Impact of array architecture on coverage patterns at 6.7~GHz: (a)--(c) radiomaps for $8\times8$, $16\times16$, and $32\times32$ UPA configurations; (d) peak-to-mean power ratio quantifying beam concentration; and (e) top-5\% power concentration measuring spatial energy focusing.}
\label{fig:array_comparison}
\end{figure*}

\begin{figure*}[t]
\centering
\includegraphics[width=0.85\linewidth]{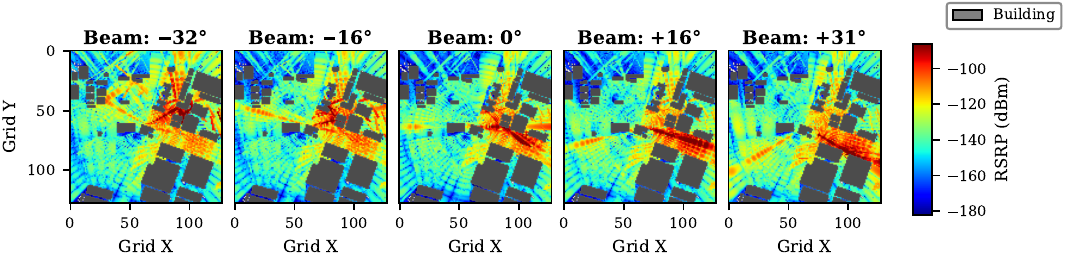}
\caption{Beam steering characteristics for a $32\times32$ UPA at 6.7~GHz across five azimuth angles ($-32^\circ$ to $+31^\circ$).}
\label{fig:beam_steering}
\end{figure*}

\textbf{Ray Tracing Configuration.} The simulation employs a maximum interaction depth of three (comprising up to three reflections or diffractions per path), $10^6$ rays per transmitter-receiver pair, and support for LoS, specular reflection, and edge diffraction propagation mechanisms. Diffuse scattering is not enabled, as preliminary experiments indicated negligible impact on path loss prediction accuracy, consistent with findings in \cite{xia2024path}.

{\textbf{Channel Matrix Computation.} For each transmitter-receiver pair and array configuration, Sionna computes the multipath channel matrix $\mathbf{H}_k \in \mathbb{C}^{N \times N_p}$ following the model in Section~\ref{sec:radiomap_modeling}. Due to the large number of receiver grid points (16,384 per radiomap) and the substantial memory footprint of channel matrices for large arrays, we employ an adaptive batching strategy that partitions receivers into groups based on available GPU memory, processes each batch independently, and aggregates results. This approach enables efficient radiomap generation across the entire configuration space without exceeding hardware constraints.}

\textbf{Beamforming and Power Calculation.} The received power at grid $k$ under precoding vector $\mathbf{w}$ is computed via beamforming combination $g_k = \mathbf{w}^H \mathbf{H}_k \mathbf{1}_{N_p}$ and power calculation $P_{r,k} = |g_k|^2 \cdot P_t$, following the normalization convention in Section~\ref{sec:radiomap_modeling}. By evaluating this computation across all $K^2$ grid points, we obtain the complete radiomap $\mathbf{P}_r$ for the given scene and array configuration.

\subsection{Multi-Configuration Parameter Space and SSB Beam Design}\label{sec:config_space}

To comprehensively characterize XL-MIMO coverage while supporting diverse machine learning research objectives, the dataset systematically varies three configuration dimensions: carrier frequency, array architecture, and beam steering direction. As summarized in Table~\ref{tab:config_space}, this parameterization yields $90 + 8 = 98$ distinct configurations per scene: {90 configurations at 6.7~GHz covering dense beam sweeping across square UPAs ($32\times32$, $16\times16$, $8\times8$) together with single-beam rectangular UPAs ($16\times32$, $8\times16$), and 8 single-beam configurations across lower frequency bands (1.8--4.9~GHz) representing legacy and current network deployments.} Combined with the 800 urban scenes acquired from the Nanjing metropolitan area (Section~\ref{sec:scene_acquisition}), the complete dataset aggregates to $800 \times 98 = 78400$ radiomaps encompassing approximately 1.28 billion spatial coverage measurements.

At 6.7~GHz, dense beam sweeping enables fine-grained beam management analysis. Beam directions are uniformly sampled in azimuth with spacing chosen so that adjacent beams overlap near their half-power points, consistent with commercial 5G SSB beam sweeping. The specific beam configurations for each array size are detailed in Table~\ref{tab:config_space}.

Across lower frequency bands (1.8--4.9~GHz), array-frequency combinations reflect practical network deployments: 3.5~GHz employs $8\times8$ UPAs representative of mainstream 5G massive MIMO, whereas 1.8~GHz utilizes smaller arrays ($4\times4$, $2\times4$, $2\times2$) typical of LTE infrastructure. This selection enables the study of model transferability across different array configurations, which is essential for network planning tools that must operate across diverse deployment scenarios without retraining for each new configuration.

\subsection{Dataset Characteristics and Visualization}

To demonstrate the richness and diversity of our dataset, we present representative visualizations across multiple dimensions.

\textbf{Geographic Diversity.} Fig.~\ref{fig:scene_diversity} illustrates three representative urban scenes with distinct morphological characteristics. Building heights range from 10~m to over 300~m, representing diverse urban morphologies from sparse suburban areas with isolated buildings to dense urban centers with closely spaced high-rise structures.

\textbf{Multi-Frequency Characteristics.} Fig.~\ref{fig:freq_comparison} compares radiomaps for a fixed $8 \times 8$ UPA across four frequency bands. As frequency increases from 2.6~GHz to 6.7~GHz, coverage range systematically decreases due to elevated path loss, while spatial granularity of multipath fading patterns increases due to shorter wavelengths. This trend is quantified in Fig.~\ref{fig:freq_comparison}(e), where the coverage ratio, i.e., the percentage of locations achieving RSRP above $-120$~dBm, drops from 57\% to 31\% as frequency increases from 2.6~GHz to 6.7~GHz.

\textbf{Array Architecture Impact.} 
Fig.~\ref{fig:array_comparison} demonstrates that larger UPA configurations produce progressively more directive radiation patterns at 6.7~GHz. We quantify this spatial power concentration using two metrics shown in Fig.~\ref{fig:array_comparison}(d)--(e). The first metric, \emph{peak-to-mean power ratio (PMR)}, is defined as $\text{PMR} = 10\log_{10}\left(\max_i P_i / \bar{P}\right)$ in dB, where $P_i$ denotes the linear received power at grid $i$ and $\bar{P}$ is the spatial average over all valid (non-building) grids. This metric increases from 22.9~dB ($8\times8$) to 28.5~dB ($32\times32$), indicating enhanced beam concentration with larger arrays. The second metric, \emph{top-5\% power concentration}, is the fraction of total power captured by the highest 5\% of locations, computed as $C_{5\%} = \sum_{i=1}^{m} P_{(i)} / \sum_{i=1}^{N} P_i$, where $P_{(i)}$ is the $i$-th largest power value and $m = \lceil 0.05N \rceil$. This metric reaches 99.1\% for the $32\times32$ UPA, demonstrating that the vast majority of received energy is concentrated within a small fraction of the coverage area.

\textbf{Beam Steering Characteristics.} Fig.~\ref{fig:beam_steering} visualizes radiomaps for a $32\times32$ UPA at 6.7~GHz across five azimuth steering angles from $-32^{\circ}$ to $+31^{\circ}$. The coverage footprint shifts correspondingly with beam direction while maintaining the narrow beamwidth characteristic of large-scale arrays, demonstrating the dataset's capability to capture beam-specific coverage patterns essential for SSB beam management research.


\begin{table*}[t]
\centering
\caption{Task Taxonomy for XL-MIMO Radiomap Prediction}
\label{tab:task_taxonomy}
\begin{tabular}{lccccc}
\toprule
\multirow{2}{*}{\textbf{Task}} & \multicolumn{3}{c}{\textbf{Input Features}} & \multirow{2}{*}{\textbf{Data Split}} & \multirow{2}{*}{\textbf{Evaluation Goal}} \\
\cmidrule(lr){2-4}
 & $\mathbf{X}_{\text{env}}$ & $\mathbf{X}_{\text{array}}$ & $\mathbf{X}_{\text{sample}}$ & & \\
\midrule
Task 1 (Blind) & \checkmark & \checkmark & $\times$ & Random & In-distribution \\
Task 2 (Sparse) & $\circ$ & $\circ$ & \checkmark & Random & In-distribution \\
Task 3a (Cross-Config) & $\circ$ & $\circ$ & $\circ$ & Config-disjoint & Extrapolation \\
Task 3b (Cross-Env) & $\circ$ & $\circ$ & $\circ$ & Scene-disjoint & Extrapolation \\
\bottomrule
\multicolumn{6}{l}{\footnotesize \checkmark: required; $\circ$: optional; $\times$: not used}
\end{tabular}
\end{table*}

In summary, the dataset comprises 78400 radiomaps spanning 800 diverse urban scenes, five frequency bands, nine array architectures, and up to 64 beam steering directions, providing comprehensive coverage characterization across the XL-MIMO configuration space.

\section{Radiomap Prediction Task Taxonomy}\label{sec:task_taxonomy}
Having established the dataset construction methodology in Section~\ref{sec:dataset}, we now define standardized evaluation protocols to systematically assess radiomap prediction methods. We first present a unified prediction framework that formulates radiomap prediction as a general mapping from available inputs to spatial coverage distributions (Section~\ref{sec:unified_framework}). Building upon this framework, we define three complementary tasks that address distinct practical requirements: blind prediction without radiomap measurements (Task~1, Section~\ref{sec:task1}), reconstruction from sparse measurements (Task~2, Section~\ref{sec:task2}), and cross-distribution generalization to unseen configurations or environments (Task~3, Section~\ref{sec:task3}).

\subsection{Unified Prediction Framework}
\label{sec:unified_framework}

Different practical scenarios impose distinct requirements on radiomap prediction models. Some applications require coverage estimation for planned deployments before any field data is available; others can leverage sparse measurements collected through drive tests or crowdsourcing; furthermore, deployed models must often generalize to array configurations or environments absent from training data.

To systematically address these diverse requirements, we formulate radiomap prediction as a unified mapping problem. Following the notation established in Section~\ref{sec:system_model}, our objective is to identify an accurate mapping function $\mathcal{G}(\cdot)$ that transforms available input features into complete radiomap predictions:
\begin{equation}
\widehat{\mathbf{P}}_r = \mathcal{G}(\mathbf{X}_{\text{array}}, \mathbf{X}_{\text{env}}, \mathbf{X}_{\text{sample}}),
\end{equation}
where $\mathbf{X}_{\text{array}}$ denotes array configuration parameters, $\mathbf{X}_{\text{env}}$ represents environmental features, $\mathbf{X}_{\text{sample}}$ corresponds to sparse measurements, and $\widehat{\mathbf{P}}_r$ is the predicted radiomap. Depending on the application scenario and data availability, each input may be required, optional, or unavailable.

Previous work typically distinguishes prediction tasks only by whether sparse measurements are available, ignoring generalization requirements and potentially leading to overly optimistic performance estimates. We therefore define tasks along two orthogonal dimensions: (1) \emph{information availability}, specifying which inputs are accessible, and (2) \emph{generalization requirement}, characterizing the distribution shift between training and testing data. Table~\ref{tab:task_taxonomy} summarizes the input-output configurations and data partitioning schemes for all tasks, distinguishing interpolation within the training distribution from extrapolation to unseen configurations or environments.

\subsection{Task 1: Blind Radiomap Prediction}\label{sec:task1}

Task~1 evaluates the model's ability to predict complete radiomap coverage from array parameters and environmental features alone, without any sparse measurements from the target area. This task corresponds to practical scenarios where coverage must be estimated for planned deployments before any field data is available.

Each data instance comprises array configuration parameters $\mathbf{X}_{\text{array}}$ (carrier frequency $f_c$, array dimensions $N_{\text{rows}} \times N_{\text{cols}}$, beam steering angles $\theta_{\text{beam}}, \phi_{\text{beam}}$) and a building height matrix $\mathbf{X}_{\text{env}}$ encoding the urban geometry as inputs, with the corresponding ground-truth radiomap $\mathbf{P}_r$ serving as the prediction target. The dataset is partitioned randomly into training and testing subsets, ensuring no data leakage while permitting potential feature space overlap (e.g., similar array configurations may appear across both subsets in different environments).

This task tests whether the model can learn the underlying propagation physics (beamforming gains, multipath interactions, shadowing effects) sufficiently well to generalize to new array-environment combinations drawn from the same distribution.

\subsection{Task 2: Sparse Measurement-Based Reconstruction}\label{sec:task2}

Task~2 evaluates the model's ability to reconstruct complete radiomaps from limited spatial measurements combined with auxiliary information. This task corresponds to practical scenarios where partial coverage data can be collected through crowdsourced user equipment reports or dedicated drive test campaigns.

Each data instance comprises sparse measurements $\mathbf{X}_{\text{sample}}$, obtained through controlled sampling (either uniform random or spatially imbalanced) from the complete radiomap depending on the application scenario, as the primary input. Array configuration parameters $\mathbf{X}_{\text{array}}$ and building height matrix $\mathbf{X}_{\text{env}}$ serve as optional auxiliary inputs, with the complete radiomap $\mathbf{P}_r$ as the prediction target. The dataset is partitioned randomly into training and testing subsets following the same protocol as Task~1.

This task tests whether the model can effectively integrate sparse measurements with prior knowledge to complete coverage maps, reflecting operational scenarios where limited field data must be extended for network optimization.

\subsection{Task 3: Cross-Distribution Generalization}\label{sec:task3}

While Tasks~1 and 2 ensure no data leakage through disjoint training-testing splits, they permit feature space overlap that may artificially inflate generalization performance estimates. When array configurations and environments vary independently across training samples, a model can learn separable representations and recombine them at test time. For example, a model trained on $32 \times 32$ arrays in environment $A$ and $16 \times 16$ arrays in environment $B$ may predict coverage for a $32 \times 32$ array in environment $B$ by composing independently learned features, rather than by truly extrapolating.

Task~3 addresses this limitation by enforcing non-overlapping feature spaces between training and testing distributions, thereby rigorously evaluating extrapolation capability beyond the training manifold. We consider two practically important generalization scenarios:
\begin{itemize}
\item \textbf{Task 3a (Cross-Configuration):} Training and testing sets contain disjoint transmitter configurations, evaluating model transferability to unseen carrier frequencies, element patterns, array architectures, or beam directions. For example, training may include sub-6~GHz arrays while testing evaluates $32\times32$ UPAs at 6.7~GHz.
\item \textbf{Task 3b (Cross-Environment):} Training and testing sets contain disjoint geographical scenes, evaluating model transferability to urban morphologies entirely absent from training data.
\end{itemize}

Task~3 provides the most stringent assessment of model robustness and transferability, reflecting real-world deployment conditions where network planning tools must accommodate configurations or locations absent from historical data. Superior performance on Task~3 indicates that the learned mapping has captured fundamental propagation principles rather than dataset-specific correlations.

Existing methods have primarily addressed cross-environment generalization (Task~3b) through sophisticated environmental encoding architectures~\cite{radiounet,radiogat,radiodiff}. These approaches excel at learning environment-dependent propagation effects from building geometry, enabling transfer to unseen urban scenes. In contrast, cross-configuration generalization (Task~3a) remains underexplored. While prior works have investigated individual dimensions such as cross-frequency~\cite{radiogat,unirm} and cross-beam-direction prediction~\cite{beamckm}, they lack unified representations that generalize across all configuration dimensions. The fundamental challenge lies in encoding array-dependent radiation characteristics: unlike environmental features that can be directly extracted from geometric data, array radiation depends on complex electromagnetic interactions (beamforming, phase coherence across array aperture, antenna directivity) that cannot be adequately captured through scalar parameter encoding. Section~\ref{sec:beam_map} addresses this gap by introducing a physics-informed beam map representation that explicitly encodes array-specific radiation patterns.

\begin{figure*}
    \centering
 \includegraphics[width=0.85\linewidth]{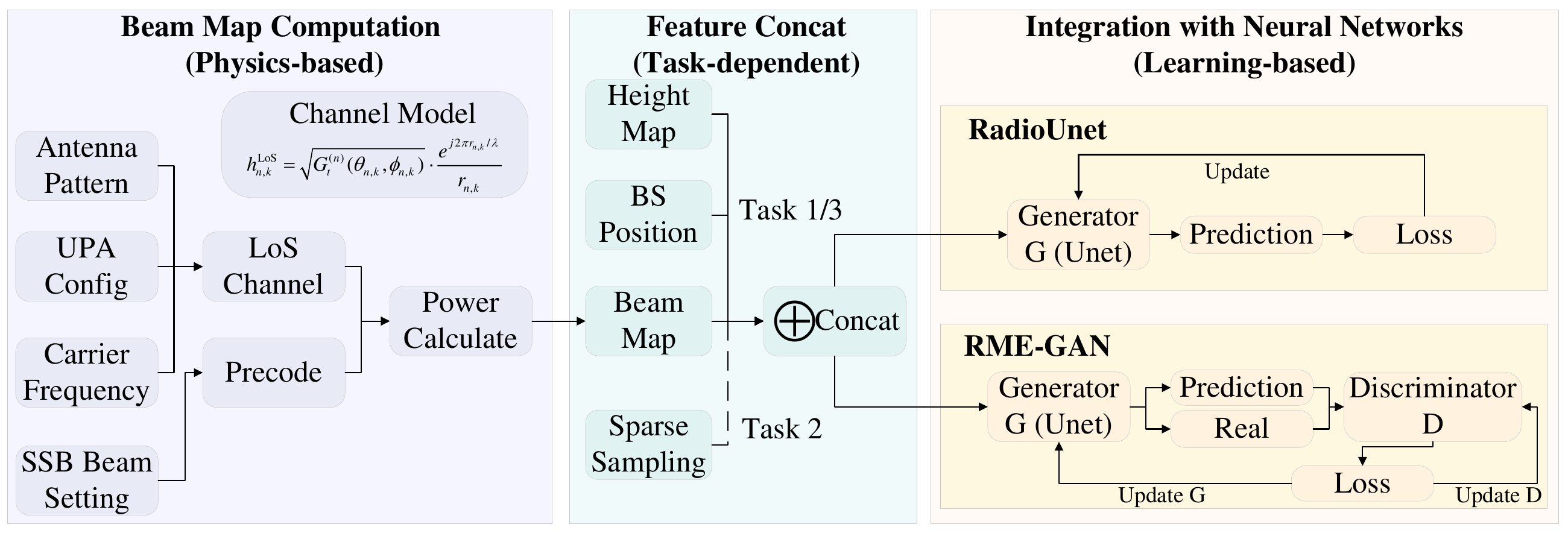}
\caption{Beam map-enhanced radiomap prediction framework. Left: physics-based beam map computation. Middle: task-dependent feature concatenation. Right: integration with neural network architectures.}
    \label{fig:beammap_framework}
\end{figure*}
\section{Beam Map Enhanced Prediction}\label{sec:beam_map}
To address the cross-configuration generalization challenge identified in Section~\ref{sec:intro}, we introduce the \emph{beam map}, a physics-informed spatial feature that provides an explicit encoding of array configuration $\mathbf{X}_{\text{array}}$ in the unified prediction framework. The key idea is to decouple radiomap prediction into two components: array-dependent radiation patterns (computed analytically via beam maps) and environment-dependent propagation effects (learned from building geometry via neural networks). By replacing implicit scalar encoding with explicit spatial encoding, this decomposition shifts the cross-configuration generalization burden from data-driven extrapolation to deterministic physical computation.

As illustrated in Fig.~\ref{fig:beammap_framework}, the proposed approach comprises: (1) physics-based beam map computation (Section~\ref{sec:beam_formulation}), and (2) integration with neural network architectures through task-dependent feature concatenation (Section~\ref{sec:beam_integration}). To validate effectiveness, we integrate beam maps into RadioUNet~\cite{radiounet} and RME-GAN~\cite{rmegan}, representing supervised and adversarial training paradigms respectively.

\subsection{Beam Map Formulation}\label{sec:beam_formulation}
The beam map is derived from the radiomap model in Section~\ref{sec:system_model} by retaining only the LoS path, for which $\Gamma_{n,p,k} = 1$ (no reflection or diffraction). Setting $N_p = 1$ in (\ref{eq3}), the channel coefficient simplifies to
\begin{equation}
h_{n,k}^{\text{LoS}} = \sqrt{G_t^{(n)}(\theta_{n,k}, \phi_{n,k})} \cdot \frac{e^{j 2\pi r_{n,k}/\lambda}}{r_{n,k}},\label{eq:beam_channel}
\end{equation}
where all terms follow definitions in Section~\ref{sec:radiomap_modeling}. The phase term $e^{j2\pi r_{n,k}/\lambda}$ with element-specific distance $r_{n,k}$ ensures accurate beamforming gain computation through coherent phase combination across the array aperture.

The LoS channel matrix becomes
\begin{equation}
\mathbf{H}_k^{\text{LoS}} = {\left[h_{1,k}^{\text{LoS}}, \ldots, h_{N,k}^{\text{LoS}} \right]}^T \in \mathbb{C}^{N \times 1}.\label{eq:beam_H}
\end{equation}
Following the power calculation procedure in (\ref{eq5})--(\ref{eq8}), the beam map value at grid $\mathcal{R}_k$ under precoding vector $\mathbf{w}$ is

\begin{equation}
B_k = \frac{\lambda^2}{(4\pi)^2} P_t \left| \mathbf{w}^H \mathbf{H}_k^{\text{LoS}} \right|^2,\label{eq:beam_value}
\end{equation}
where $P_t = 1$~mW follows the normalization convention established in Section~\ref{sec:power_normalization}. The coherent superposition $\mathbf{w}^H \mathbf{H}_k^{\text{LoS}}$ captures antenna element directivity, beamforming gain, and LoS propagation pattern. By computing $B_k$ for all grids in the observation plane, we obtain the complete beam map:

\begin{equation}
\mathbf{B} = \begin{bmatrix}
B_{1} & \cdots & B_{K} \\
\vdots & \ddots & \vdots \\
B_{K^2-K+1} & \cdots & B_{K^2}
\end{bmatrix},\label{eq:beam_map}
\end{equation}
providing a $K \times K$ spatial representation of the ideal LoS coverage pattern for the given array configuration and beam steering direction.

The beam map encodes three physical effects: (1) antenna element radiation patterns, (2) spatially varying free-space propagation loss, and (3) directional focusing from beamforming. It is computed analytically from array configuration parameters alone without environmental information. This property ensures that the beam map serves as a self-contained representation of $\mathbf{X}_{\text{array}}$, independent of the specific environment or neural network architecture.

\subsection{Integration with Deep Learning Architectures}\label{sec:beam_integration}

The beam map provides an explicit spatial encoding of $\mathbf{X}_{\text{array}}$ in the unified prediction framework established in Section~\ref{sec:unified_framework}. Recall that the mapping function $\mathcal{G}(\mathbf{X}_{\text{array}}, \mathbf{X}_{\text{env}}, \mathbf{X}_{\text{sample}})$ requires array configuration as input. Existing methods encode $\mathbf{X}_{\text{array}}$ implicitly by spatially tiling scalar parameters (frequency, array dimensions, beam angles) as input channels. This implicit encoding forces neural networks to learn array-dependent radiation patterns from low-dimensional scalar inputs, which limits generalization to unseen configurations.

In contrast, beam map-enhanced models replace this implicit encoding with the analytically computed beam map $\mathbf{B}$. Since $\mathbf{B}$ is a $K \times K$ spatial feature with the same resolution as the radiomap, it can be directly concatenated with environmental features $\mathbf{X}_{\text{env}}$ and optional sparse measurements $\mathbf{X}_{\text{sample}}$ as multi-channel inputs to the encoder. The feature concatenation is task-dependent: Task~1 excludes sparse measurements for blind prediction, Task~2 requires them for reconstruction, and Task~3 optionally incorporates them depending on the evaluation scenario (cf. Table~\ref{tab:task_taxonomy}). This integration requires no architectural modification, making the beam map representation applicable to arbitrary encoder-decoder backbones.

The key advantage of this formulation is that array-dependent radiation characteristics are now explicitly provided as spatial features, allowing the neural network to focus on learning environment-dependent propagation effects (shadowing, multipath, diffraction). This decomposition aligns with the physical structure of radiomap prediction and enables cross-configuration generalization without requiring the network to extrapolate array behaviors from scalar parameters.

\section{Experiment}\label{sec:experiment}

In this section, we conduct comprehensive experiments on the proposed XL-MIMO radiomap dataset to validate the effectiveness of the beam map-enhanced prediction paradigm. The diverse array configurations and urban scenes in the dataset enable rigorous evaluation across blind prediction, sparse reconstruction, and cross-distribution generalization tasks.

\subsection{Experimental Setup}
\textbf{Dataset and Partitioning.}
Following the protocols defined in Section~\ref{sec:task_taxonomy}, we partition the dataset with 7:1:2 ratio for training, validation, and testing along different dimensions depending on the task. For Tasks~1 and 2, all 78400 radiomap instances are randomly partitioned at the instance level. For Task~3a (cross-configuration), the 98 array-beam configurations are partitioned into 68/9/21 for training/validation/testing, while all 800 scenes appear across all three subsets; this yields 54400/7200/16800 radiomap instances respectively. For Task~3b (cross-environment), the 800 scenes are partitioned into 560/80/160, while all 98 configurations appear across all subsets; this yields 54880/7840/15680 radiomap instances respectively.

\textbf{Input Configuration.}
We use environmental features $\mathbf{X}_{\text{env}}$ (building height maps) and array configuration $\mathbf{X}_{\text{array}}$ as inputs for all tasks, where $\mathbf{X}_{\text{array}}$ is encoded either implicitly via spatially tiled parameter vectors (baseline) or explicitly via beam maps (proposed). We also include a binary mask indicating transmitter location as an auxiliary input for completeness. Task~2 additionally incorporates sparse measurements $\mathbf{X}_{\text{sample}}$ at 5\% sampling rate.

\textbf{Baselines and Metrics.}
We use RadioUNet and RME-GAN as baselines, representing encoder-decoder and adversarial training paradigms. Since the beam map is a model-agnostic input feature, consistent improvements across both architectures would demonstrate its general effectiveness. As described in Section~\ref{sec:beam_integration}, baseline models employ implicit configuration encoding via spatially tiled parameter vectors, while beam map-enhanced models replace this encoding with explicit spatial beam maps. This comparison isolates the benefit of physics-informed spatial feature representation. Performance is evaluated using MAE and Root Mean Squared Error (RMSE) in dB.

\textbf{Implementation.} We adopt the official open-source implementation of RadioUNet~\cite{radiounet}, which employs a W-Net architecture with two cascaded U-Net-style generators. Each generator uses downsampling blocks with channel dimensions [64, 128, 256], ResNet blocks at the bottleneck, and symmetric upsampling blocks. RME-GAN adopts the conditional GAN framework from~\cite{rmegan} with a PatchGAN\cite{isola2017image} discriminator operating on $70\times70$ receptive fields; the generator loss combines adversarial loss (weight 0.5) and L1 reconstruction loss (weight 400).

All models process $256 \times 256$ input tensors\footnote{The generated radiomaps ($128 \times 128$) and building height matrices ($256 \times 256$) are bilinearly interpolated to a unified $256 \times 256$ resolution to maintain compatibility with the original network architectures.}. RadioUNet training employs the Adam optimizer with learning rate $10^{-4}$ and step decay (factor 0.1 every 30 epochs), batch size 256, for 5 epochs per stage. RME-GAN uses learning rate $10^{-5}$, batch size 128, for 20 epochs with best model selection based on validation loss. Beam maps are pre-computed offline via (\ref{eq:beam_channel})--(\ref{eq:beam_map}) in under 10~ms per configuration.

\subsection{Main Results}

\begin{table}[t]
\centering
\caption{Radiomap Prediction Performance: Blind Prediction (Task 1) vs. Sparse Reconstruction at 5\% Sampling (Task 2)}
\label{tab:task1_task2_results}
\small
\setlength{\tabcolsep}{4pt}
\begin{tabular}{llccc}
\toprule
\textbf{Task} & \textbf{Method} & \textbf{MAE (dB)} & \textbf{RMSE (dB)} & \textbf{$\Delta$MAE} \\
\midrule
\multirow{4}{*}{\begin{tabular}[c]{@{}l@{}}Task 1\\(Blind)\end{tabular}} 
 & RadioUNet & 12.98 & 16.98 & -- \\
 & RadioUNet+Beam & 7.69 & 11.61 & $\downarrow$ 40.8\% \\
\cmidrule(lr){2-5}
 & RME-GAN & 12.94 & 17.05 & -- \\
 & RME-GAN+Beam & \textbf{6.27} & \textbf{10.53} & $\downarrow$ 51.5\% \\
\midrule
\multirow{4}{*}{\begin{tabular}[c]{@{}l@{}}Task 2\\(Sparse)\end{tabular}} 
 & RadioUNet & 8.73 & 12.03 & -- \\
 & RadioUNet+Beam & 5.87 & 8.86 & $\downarrow$ 32.8\% \\
\cmidrule(lr){2-5}
 & RME-GAN & 8.01 & 11.59 & -- \\
 & RME-GAN+Beam & \textbf{5.56} & \textbf{8.61} & $\downarrow$ 30.6\% \\
\bottomrule
\end{tabular}
\end{table}

\begin{table}[t]
\centering
\caption{Cross-Distribution Generalization Performance (Task 3)}
\label{tab:task3_generalization}
\small
\setlength{\tabcolsep}{2pt}
\begin{tabular}{llccc}
\toprule
\textbf{Scenario} & \textbf{Method} & \textbf{MAE (dB)} & \textbf{RMSE (dB)} & \textbf{$\Delta$MAE} \\
\midrule
\multirow{4}{*}{\begin{tabular}[c]{@{}l@{}}Cross-Config\\(Task 3a)\end{tabular}} 
 & RadioUNet & 19.91 & 25.55 & -- \\
 & RadioUNet+Beam & 7.96 & 12.06 & $\downarrow$ 60.0\% \\
\cmidrule(lr){2-5}
 & RME-GAN & 13.27 & 17.41 & -- \\
 & RME-GAN+Beam & \textbf{6.24} & \textbf{10.38} & $\downarrow$ 53.0\% \\
\midrule
\multirow{4}{*}{\begin{tabular}[c]{@{}l@{}}Cross-Env\\(Task 3b)\end{tabular}} 
 & RadioUNet & 13.16 & 17.23 & -- \\
 & RadioUNet+Beam & 7.85 & 12.12 & $\downarrow$ 40.4\% \\
\cmidrule(lr){2-5}
 & RME-GAN & 12.76 & 16.79 & -- \\
 & RME-GAN+Beam & \textbf{6.31} & \textbf{10.66} & $\downarrow$ 50.5\% \\
\bottomrule
\end{tabular}
\end{table}

\subsubsection{Task 1 \& Task 2: Blind Prediction and Sparse Reconstruction}
We first examine whether the proposed beam map representation improves radiomap prediction accuracy. Table~\ref{tab:task1_task2_results} presents quantitative results for blind radiomap prediction and sparse reconstruction at 5\% sampling rate. Beam map integration improves both architectures on both tasks. The improvement is larger for Task~1 (40.8--51.5\% MAE reduction) than Task~2 (30.6--32.8\%), suggesting that beam maps are most valuable when sparse measurements are unavailable. Combining beam maps with sparse measurements (Task~2) yields the best performance, indicating that the two information sources are complementary.

\begin{figure*}[t]
    \centering
    \includegraphics[width=0.84\linewidth]{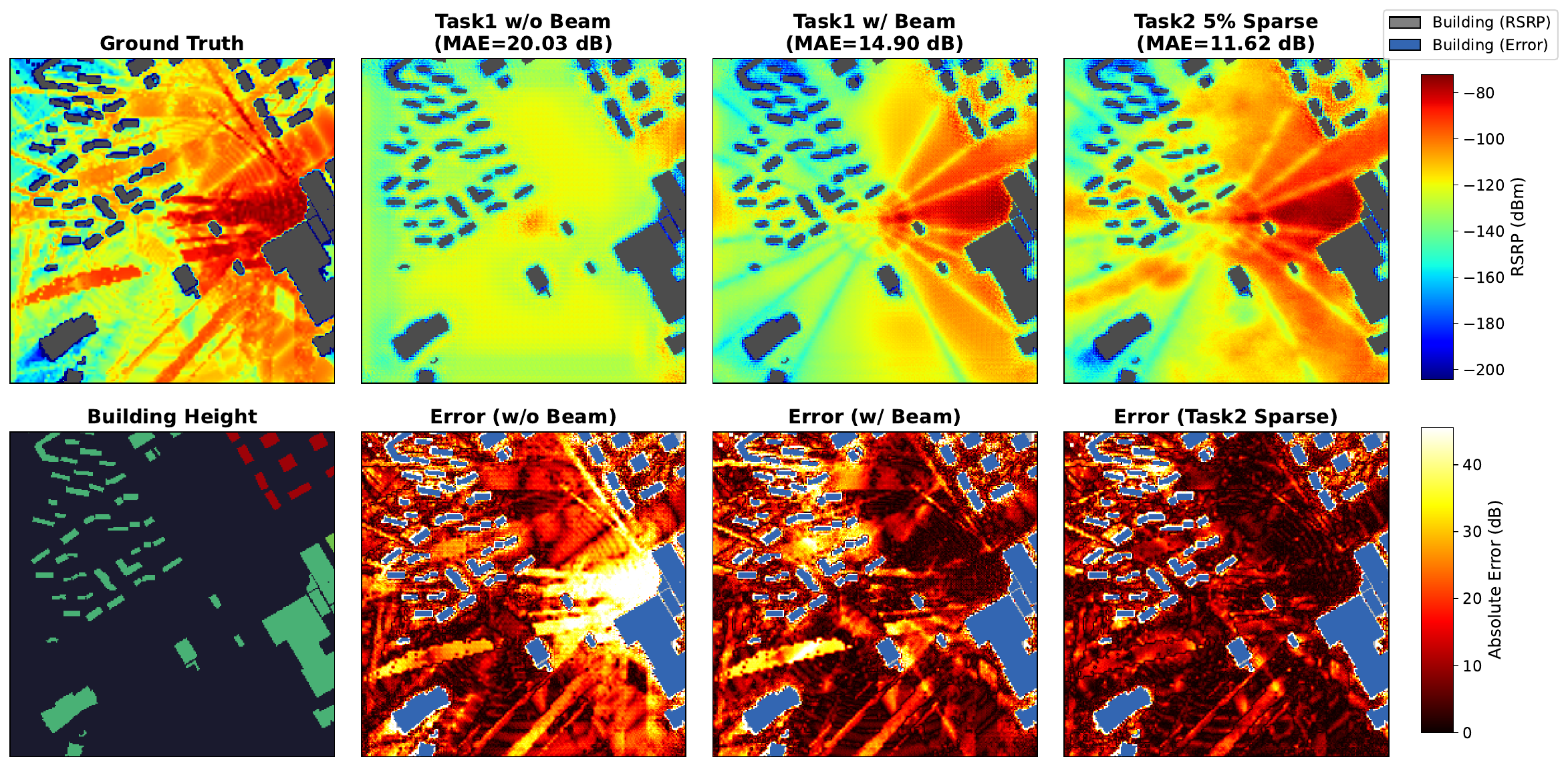}
\caption{Qualitative comparison of radiomap prediction methods on a representative example. Top: ground-truth and predictions. Bottom: building heights and absolute error distributions.}
    \label{fig:diffraction_comparison}
\end{figure*}

Fig.~\ref{fig:diffraction_comparison} illustrates prediction quality on a selected example. Without beam maps, the baseline (MAE=20.03~dB) produces smoothed predictions that lack directional structure. With beam maps (MAE=14.90~dB), the prediction captures the main lobe and coverage boundaries. Adding sparse measurements (Task~2, MAE=11.62~dB) further improves accuracy in diffraction regions.

\subsubsection{Task 3: Cross-Distribution Generalization}
We next investigate whether beam maps enhance model generalization to unseen array configurations and environments. Table~\ref{tab:task3_generalization} presents model performance on feature distributions entirely unseen during training. While direct comparison with Task~1 is not strictly valid due to different partitioning strategies, the substantial performance degradation from 12.98~dB (Task~1) to 19.91~dB (Task~3a) and 13.16~dB (Task~3b) for baseline RadioUNet indicates that cross-distribution generalization is inherently more challenging than in-distribution prediction.

The baseline results further reveal that cross-configuration generalization (Task~3a) is harder than cross-environment generalization (Task~3b): RadioUNet's MAE is over 50\% worse on Task~3a. This suggests that neural networks can transfer environmental patterns to new scenes but struggle to generalize array-specific behaviors from scalar parameter encodings.

Beam map integration largely closes this gap. RadioUNet+Beam achieves 7.96~dB on Task~3a and 7.85~dB on Task~3b, demonstrating near-parity across both generalization scenarios. The improvement is most pronounced for cross-configuration generalization: 60.0\% MAE reduction on Task~3a versus 40.4\% on Task~3b. RME-GAN shows consistent trends (53.0\% and 50.5\% reductions, respectively).

The distinct improvement magnitudes reflect the different mechanisms through which beam maps enhance generalization. For cross-environment generalization (Task~3b), the improvement stems from enhanced representational capacity: beam maps provide richer spatial features that help the network learn more robust environment-propagation mappings, similar to the gains observed in Task~1. For cross-configuration generalization (Task~3a), the improvement is more fundamental: since beam maps are computed analytically from array parameters, they inherently provide accurate radiation patterns for any configuration, eliminating the need for neural networks to extrapolate array-specific behaviors from scalar encodings. This explains why beam maps nearly equalize performance across the two generalization scenarios.

\begin{figure}[t]
    \centering
    \includegraphics[width=0.95\linewidth]{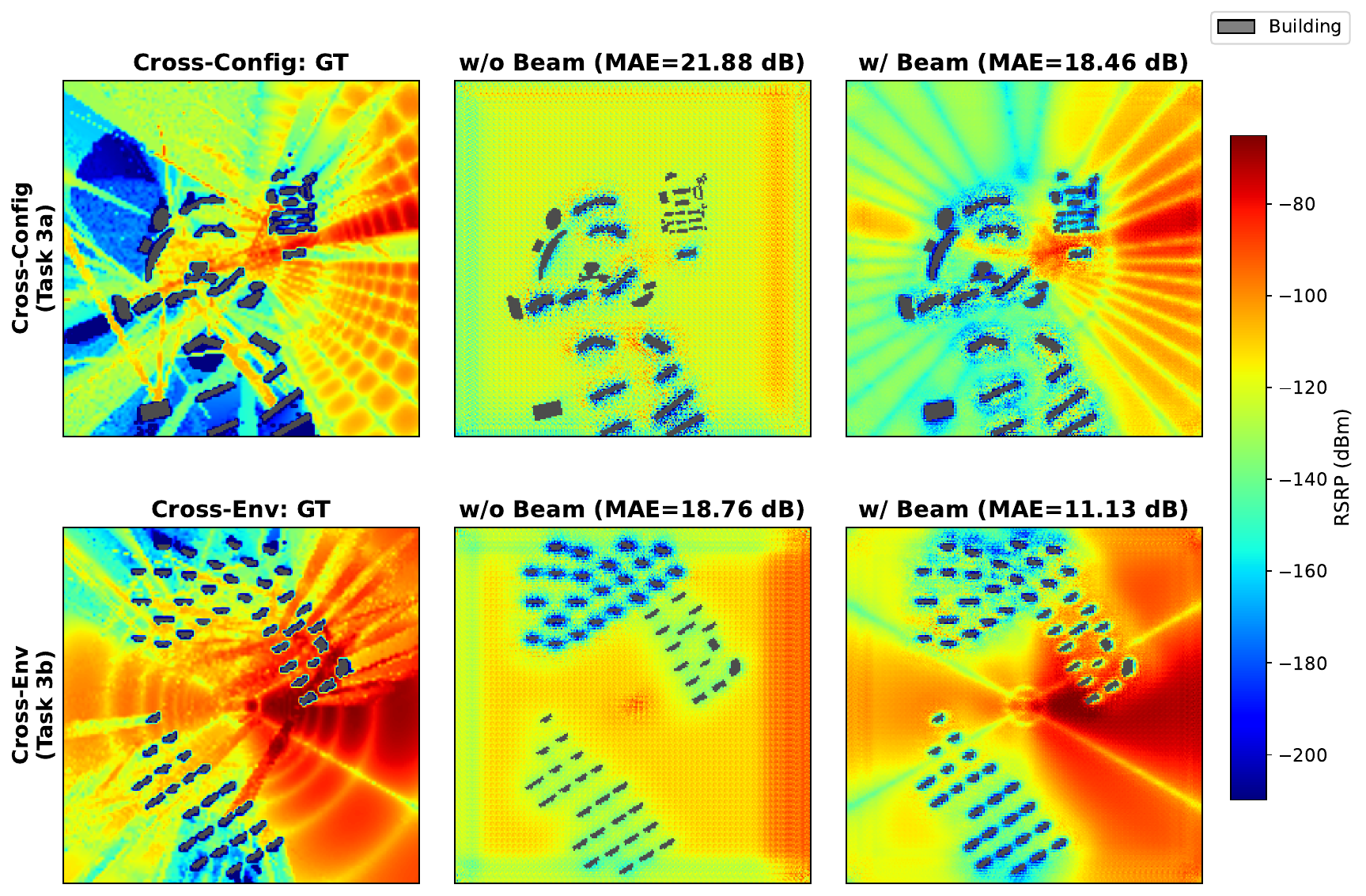}
\caption{Cross-distribution generalization comparison. Top: Task 3a (cross-configuration). Bottom: Task 3b (cross-environment). Each row: ground truth, baseline prediction, and beam map-enhanced prediction.}
    \label{fig:generalization_comparison}
\end{figure}

Fig.~\ref{fig:generalization_comparison} visualizes cross-distribution generalization. In the cross-configuration scenario (top row), the baseline produces incorrect beam direction and diffuse spatial structure, whereas beam map integration accurately captures the directional pattern of the unseen $16\times16$ UPA. In the cross-environment scenario (bottom row), beam maps yield sharper coverage boundaries and reduced error in shadowed regions.

\subsection{Discussion on Practical Significance}
The proposed beam map is a physics-informed, model-agnostic representation for cross-configuration generalization rather than a per-scenario over-optimized predictor. Accordingly, the reported 6~dB MAE should be interpreted under strict cross-distribution settings rather than easier random splits. Comparable error levels have also been reported under practical prediction or strict generalization settings \cite{11268973,bakirtzis2025radiopropagationmodellingdifferentiate,unirm}. In particular, \cite{11268973} reports 5~dB-level MAE in practical low-altitude coverage prediction and 6.26~dB MAE with demonstrated transferability, while \cite{bakirtzis2025radiopropagationmodellingdifferentiate} reports comparable MAE levels in large-scale real-network radio propagation modeling. In RMSE terms, \cite{unirm} reports 20.844~dB at 1.5~m height, whereas our beam map-enhanced models achieve roughly 10~dB RMSE. These results support the practical relevance of our method and highlight that beam maps reduce error under configuration shift while providing a transferable representation for future predictors.

\begin{table}[t]
\centering
\caption{Ablation Study on Input Feature Configurations (Task 1, RadioUNet)}
\label{tab:ablation_features}
\small
\setlength{\tabcolsep}{4pt}
\begin{tabular}{lcc}
\toprule
\textbf{Input Configuration} & \textbf{MAE (dB)} & \textbf{RMSE (dB)} \\
\midrule
Environment Only & 13.52 & 17.41 \\
Env + Config Encoding & 12.98 & 16.98 \\
Env + Beam Map & \textbf{7.69} & \textbf{11.61} \\
\bottomrule
\end{tabular}
\end{table}

\subsection{Ablation Study}

We further examine the contribution of beam map computation and over implicit parameter encoding. Table~\ref{tab:ablation_features} compares different input configurations using RadioUNet on Task~1. The results reveal a clear hierarchy: implicit parameter encoding provides only marginal benefit (reducing MAE from 13.52~dB to 12.98~dB), while explicit spatial encoding via beam maps yields substantial gains (7.69~dB), validating the importance of physics-informed array radiation encoding for XL-MIMO radiomap prediction.

\section{Conclusion}\label{sec:conclusion}
This paper addressed the radiomap prediction problem for XL-MIMO systems through three contributions: a large-scale dataset supporting arrays up to $32\times32$ UPAs across diverse configurations, a systematic evaluation framework enabling standardized benchmarking, and the beam map representation that analytically encodes array-specific radiation patterns. The key insight is that array-dependent line-of-sight coverage can be computed deterministically rather than learned, thereby shifting generalization from data-driven extrapolation to physics-based computation. Experimental results validated consistent improvements across architectures and tasks, with particularly substantial gains when predicting radiomaps for array configurations absent from training data. The dataset and benchmarks provide essential infrastructure for future research on beam-aware prediction and AI-driven network planning for next-generation wireless systems.

To ensure full reproducibility, we publicly release the complete dataset, including radiomaps, height maps, beam maps, configuration files, ray-tracing scenes, and simulation parameters, together with pretrained models and the end-to-end code for data generation, training, and evaluation, at {https://lxj321.github.io/MulticonfigRadiomapDataset/}.


%





\ifCLASSOPTIONcaptionsoff
  \newpage
\fi



%
\ifCLASSOPTIONcaptionsoff
  \newpage
\fi

\bibliographystyle{IEEEtran}
\bibliography{bibtex/bib/IEEEexample}




%







\end{document}